\documentclass[12pt]{article}
\usepackage{mathrsfs}
\usepackage{amssymb}
\normalsize
\input{epsf}

\begin{document}
\addtolength{\baselineskip}{.20mm}
\newlength{\extraspace}
\setlength{\extraspace}{2mm}
\newlength{\extraspaces}
\setlength{\extraspaces}{2mm}

\newcommand{\newsection}[1]{
\vspace{15mm} \pagebreak[3] \addtocounter{section}{1}
\setcounter{subsection}{0} \setcounter{footnote}{0}
\noindent {\Large\bf \thesection. #1} \nopagebreak
\medskip
\nopagebreak}

\newcommand{\newsubsection}[1]{
\vspace{1cm} \pagebreak[3] \addtocounter{subsection}{1}
\addcontentsline{toc}{subsection}{\protect
\numberline{\arabic{section}.\arabic{subsection}}{#1}}
\noindent{\large\bf 
\thesubsection. #1} \nopagebreak \vspace{3mm} \nopagebreak}
\newcommand{\ba}{\begin{eqnarray}
\addtolength{\abovedisplayskip}{\extraspaces}
\addtolength{\belowdisplayskip}{\extraspaces}

\addtolength{\belowdisplayshortskip}{\extraspace}}

\newcommand{\be}{\begin{equation}
\addtolength{\abovedisplayskip}{\extraspaces}
\addtolength{\belowdisplayskip}{\extraspaces}
\addtolength{\abovedisplayshortskip}{\extraspace}
\addtolength{\belowdisplayshortskip}{\extraspace}}
\newcommand{\ee}{\end{equation}}
\newcommand{\STr}{{\rm STr}}
\newcommand{\figuur}[3]{
\begin{figure}[t]\begin{center}
\leavevmode\hbox{\epsfxsize=#2 \epsffile{#1.eps}}\\[3mm]
\parbox{15.5cm}{\small
\it #3}
\end{center}
\end{figure}}
\newcommand{\im}{{\rm Im}}
\newcommand{\calm}{{\cal M}}
\newcommand{\call}{{\cal L}}
\newcommand{\sect}[1]{\section{#1}}
\newcommand\hi{{\rm i}}
\def\bea{\begin{eqnarray}}
\def\eea{\end{eqnarray}}

\begin{titlepage}
\begin{center}

\vspace{3.5cm}

{\Large \bf{The Slow-Roll and Rapid-Roll Conditions  in The Space-like Vector Field Scenario}}\\[1.5cm]

{Yi Zhang $^{a,b,}$\footnote{Email: zhangyi@itp.ac.cn},}
\vspace*{0.5cm}

{\it $^{a}$ Institute of Theoretical Physics, \\Chinese Academy of
Sciences P.O.Box 2735, Beijing 100190, China

$^{b}$Graduate University of Chinese Academy of Sciences,
\\ YuQuan
Road 19A, Beijing 100049, China}

\date{\today}
\vspace{3.5cm}

\textbf{Abstract} \vspace{5mm}

\end{center}
In this note we derive the slow-roll and rapid-roll conditions for
the minimally and non-minimally coupled space-like vector fields.
The  function $f(B^{2})$  represents  the non-minimal coupling
effect between vector fields and gravity, the $f=0$ case is the
minimal coupling case. For a clear comparison with scalar field,
 we define a new function $F=\pm B^{2}/12+f(B^{2})$ where
$B^{2}=A_{\mu}A^{\mu}$, $A_{\mu}$ is the ``comoving" vector field.
 With reference to
the slow-roll and rapid-roll conditions, we find the small-field
model is more suitable than the large-field model in the minimally
coupled vector field case. And  as a non-minimal coupling example,
the $F=0$ case just has the same slow-roll conditions as the scalar
fields.
\end{titlepage}

\section{Introduction}\label{sec1}
  Inflation, as a theory to describe the period of acceleration in the early
  universe,
was introduced as a way to solve the problems in the standard
big-bang theory \cite{Guth:1980zm,Linde:1981mu}.   Dark energy, as a
candidate for the period of acceleration in the late universe, was
suggested by a combination of different cosmic probes that primarily
involves Supernova data \cite{Riess:1998cb,Perlmutter:1998np}. The
scalar field is the most popular candidate for the dynamical source
of the two accelerations. We could
  assume the scalar field isotropic and
homogeneous naturally. The assumption that the scalar field is
homogeneous and isotropic coincides with the observable isotropic
and homogeneous Friedmann-Robertson-Walker (FRW) background
automatically. However, the fundamental scalar
 field, has not been probed until now. On the contrary,
 the vector field is common in the realistic world. The vector
fields inflation scenario was proposed by Ref.
\cite{Ford:1988wq,Ford:1989me}, recently extended to higher spin
field \cite{Germani:2009iq,Kobayashi:2009hj,Koivisto:2009sd}.

 The
vector field $A_{\mu}=(A_{0},A_{1},A_{2},A_{3})$ is anisotropic in
nature for the oriented components. To coincide with the observable
isotropic and homogenous FRW background, there are three models
which give out isotropy in vector field scenarios. The first one is
to the vector fields  $A_{\mu}=(A_{0},0,0,0)$ with only the temporal
component which is isotropic obviously
\cite{Kiselev:2004py,Carroll:2004ai, Lim:2004js,Boehmer:2007qa,
Koivisto:2008xf,Koh:2009vm,Koh:2009ne,Carroll:2009em,Li:2007vz}. The
second model, called ``cosmic triad", has  three spatial components
 equal to each other and orthogonal to each other in which the vector
field has such a form $A_{\mu}=(0,A,A,A)$
\cite{Bento:1992wy,ArmendarizPicon:2004pm,Wei:2006tn}£¬(see also
\cite{Hosotani:1984wj,Galtsov:1991un,Zhao:2005bu,Zhao:2006mk,Bamba:2008ja,Bamba:2008xa}
for  exact isotropic solutions of the Einstein-Yang-Mills system
based on the same idea). And the third scenario, called ``N-flation"
vector scenario, has a large number of randomly oriented fields
\cite{Golovnev:2008cf,Golovnev:2008hv,Golovnev:2009ks} in which the
vector field has a form as $A_{\mu}=(0,A_{1},A_{2},A_{3})$. Under
certain approximations, the
 forms of last two space-like scenarios are very similar
to each other, which play the leading role in this note.

However, even after the isotropic problem is solved, the slow-roll
problem needs to be solved to  make the duration of inflation last
long enough in  vector field scenarios. There are two questions
about the slow-roll problem:  one is what's the exact form of the
slow-roll conditions in the vector field scenarios; the other is
whether the de-Sitter phase will appear or not without slow rolling
vector field. Whatever, the dark energy dominating acceleration
phase only requires one e-folding number by observations. Even the
vector field driven acceleration could not offer  the e-folding
number as large as $60$ for inflation,  a period of acceleration
before or after the  main part of inflation can also alleviate some
cosmic problems (such as the moduli problem) \cite{Linde:2001ae}. It
seems that  not only the slow-roll conditions, but also the
rapid-roll conditions which is looser proposed by Ref.
\cite{Kofman:2007tr,Chiba:2008ia,Chiba:2008rp} are worthy of
considering. Because under rapid-roll conditions the universe can
get a de-Sitter phase as well. Moreover, based on  the non-minimal
coupling effect of changing  dynamics of  vector field, we will
consider the non-minimal coupling term in the discussions.

In the following section, a concise introduction will be given out
for ``cosmic triad" and ``$N$-flation" vector field scenarios. They
can be expressed in a similar form  with the non-minimally coupled
scalar field. Then,  the non-minimally coupled vector field scenario
is given out as well. In Sec. \ref{3} and \ref{4}, the slow-roll and
rapid-roll conditions
 both in the minimal and non-minimal coupling cases will be discussed as our main aim.
 Special
 examples will be given out during the descriptions.

\section{ Space-like Vector Field Scenarios}\label{sec2}
 To make a complete
description, it is appropriate to consider the possibility of
$w=p/\rho<-1$ \cite{ArmendarizPicon:2004pm,ArmendarizPicon:2003qw}
which is  suggested by dark energy observations \cite{Riess:2004nr}.
The
 discussions on  both the positive  and  negative
 kinetic energy
 cases in the vector field scenario are included in the note.
  Although the negative kinetic energy case may  have a lot of
 theoretical problems, it may be phenomenologically significant
 and worth putting other theoretical difficulties aside temporally.

\subsection{ ``Cosmic Triad" Vector Field Scenario}\label{subsec2.1}
The ``cosmic triad" vector field scenario
\cite{ArmendarizPicon:2004pm}  composes by a set of three identical
self-interacting vectors which could naturally arise (for instance
from a gauge theory with $SU(2)$ or $SO(3)$ gauge group). The three
vector fields, which are minimally coupled with gravity, have the
action
  \be
  \label{action1}
  S=\int d^{4}x \sqrt{-g}\left[\frac{R}{16 \pi G}-\sum^{3}_{a=1}\left(\pm\frac{1}{4}F^{a}_{\mu\nu}F^{a\mu\nu}+V(A^{a2})\right)\right],
  \ee
where
$F^{a}_{\mu\nu}=\partial_{\mu}A_{\nu}^{a}-\partial_{\nu}A_{\mu}^{a}$
and $A^{a2}=g^{\mu\nu}A_{\mu}^{a}A_{\nu}^{a}$,  Latin indices label
the different fields ($a,b,...=1...3$), and Greek indices label the
different space-time components ($\mu,\nu,...=0...3$). The term $\pm
F^{a}_{\mu\nu}F^{a\mu\nu}/4$ could be considered as the Maxwell type
kinetic energy term, the case with a sign $``+"$ corresponds to the
positive kinetic energy term and the case with a sign $``-"$
corresponds to the negative kinetic energy case in the vector field
scenario. The term $V(A^{2})$ is the potential of the vector field.
 The four-dimension homogenous and
isotropic FRW metric   has such a form
 \be
  ds^{2}=-dt^{2}+a^{2}(t)\delta_{ik}dx^{i}dx^{k},
 \ee
 where $a$ is the scale factor,  we consider the spatial curvature $k=0$.
And we could define  a new variable called ``physical" vector field
$B_{i}$ as discussed in Ref. \cite{ArmendarizPicon:2004pm}
 \be
  \label{B}
  B_{i}=\frac{A_{i}}{a}=aA^{i},
  \ee
 where $ A_{i}$ is called ``comoving" vector field.
The related equation $B^{2}=B_{i}B_{i}=A_{\mu}A^{\mu}=A^{2}$ could
be  conveniently gotten in the FRW background. Then we will express
most equations in term of $B_{i}$ and $B^{2}$ in the following
discussions.

By varying the action in Eq. (\ref{action1}) with respect to
$g_{\mu\nu}$, the $(00)$ and $(ii)$ components in the Einstein
equations,   called Friedmann  and Raychaudhuri equation, can
 be obtained
\begin{eqnarray}
  \label{frw1}
 &&H^{2}=\frac{8\pi G }{3}\rho,\\
 &&\dot{H}=-4\pi G(\rho+p), \label{frw2}
   \end{eqnarray}
where $H=\dot{a}/a$ is the Hubble parameter. The ansatz in ``cosmic
triad", that the three vectors are equal and orthogonal to each
other, can be expressed as
  \be
  A^{b}_{\mu}=\delta^{b}_{\mu}B(t)\cdot a.
 \ee
  The corresponding
energy density $\rho$ and pressure $p$ are given by
\begin{eqnarray}
 &&\rho=\pm\frac{3}{2}(\dot{B}_{i}+H B_{i})^{2}+3V(B^{2}),\\
 &&p=\pm\frac{1}{2}(\dot{B}_{i}+H B_{i})^{2}-3V(B^{2})+V'_{i}B_{i},
  \end{eqnarray}
where the dot means a derivative with respect to time $t$, and the
prime  with an index $i$ denote a derivative with respect to vector
field $B_{i}$, for example $V_{i}'=dV/dB_{i}$,   two primes used in
the following calculations are given by
$''=d^{2}/dB_{i}^{2}=d^{2}/dB^{2}$. And equations of  motion of
vector field could be obtained by varying the action in Eq.
(\ref{action1}) with respect to vector fields $A^{a}_{\mu}$
 \be
  \label{me1}
  \ddot{B}_{i}+3H\dot{B}_{i}+(2H^{2}+\dot{H})B_{i}\pm V'_{i}=0.
 \ee
 In the above equations, even when $V=constant$, the
term $(2H^{2}+\dot{H})B_{i}$ can make  $B_{i}$ evolve as if there
were an additional effective potential $(2H^{2}+\dot{H})B^{2}/2$.

For a clear comparison with scalar field, we define a function
$F(B^{2})=\pm B^{2}/12$. The  forms of  energy density,
pressure and  equations of motion in  vector field become
\begin{eqnarray}
 \label{rho}
 &&\rho=3\left(\pm\frac{1}{2}\dot{B}_{i}^{2}+V(B^{2})+6H(\dot{F}+H F)\right),\\
   \label{p}
 &&p=3\left(\pm\frac{1}{2}\dot{B}_{i}^{2}-V(B^{2})-2\ddot{F}-4H\dot{F}-2F(2\dot{H}+3H^{2})\right),\\
 \label{m}
  &&\ddot{B}_{i}+3H\dot{B}_{i}\pm6F'_{i}(2H^{2}+\dot{H})\pm V'_{i}=0,
\end{eqnarray}
 which
  are very similar to those in the non-minimally coupled scalar
 field. The latter has a action like
 \be
 \label{action}
 S=\int d^{4}x \sqrt{-g}\left(\frac{R}{16 \pi G}+\frac{(\nabla\phi)^{2}}{2}- V(\phi)-
 f(\phi)R\right),
 \ee
 where $f(\phi)$ denotes the non-minimal coupling  between the  field
 and  gravity. Then the forms of energy density,   pressure and  equation
 of motion for scalar field are
 \begin{eqnarray}
 \label{rhos}
 &&\rho=\frac{1}{2}\dot{\phi}^{2}+V(\phi^{2})+6H(\dot{f}+H f),\\
  \label{ps}
  &&p=\frac{1}{2}\dot{\phi}^{2}-V(\phi^{2})-2\ddot{f}-4H\dot{f}-2f(2\dot{H}+2H^{2}),\\
  \label{ms}
   &&\ddot{\phi} +3H\dot{\phi}+V'+6f'(\dot{H}+2H^{2})=0.
   \end{eqnarray}
where the prime in Eq. (\ref{ms}) is a derivative with
  respect to  scalar field $\phi$.
The differences between  Eq. (\ref{rho}) and Eq. (\ref{rhos}), and
Eq. (\ref{p}) and Eq. (\ref{ps}) are only in the coefficients. And
the equations of motion   are nearly the same.

\subsection{``N-flation" Vector Field Scenario}\label{subsec2.2}
The ``N-flation" vector field scenario \cite{Golovnev:2008cf},
inspired by the ``N-flation" scalar field model
\cite{Dimopoulos:2005ac}, has $N$ randomly oriented vector fields.
Following the assumptions in Ref. \cite{Golovnev:2008cf},
 all the vector fields have equal potentials and  same orders
of initial values. In ``N-flation" vector field scenario, although
the  anisotropy can be counterbalanced by the randomly oriented
fields mainly, the universe is slightly anisotropic. Concretely
speaking, until the end of inflation, the vector fields will remain
an anisotropy of order $1/\sqrt{N}$. As long as $N$ is large enough,
the anisotropy could be ignored \cite{Golovnev:2008cf}. The equation
 \be
 \sum^{N}_{a=1}B_{i}^{a}B_{j}^{a}\simeq\frac{N}{3}B^{2}\delta^{i}_{j}+{\cal
 O}(1)\sqrt{N}B^{2},
 \ee
 notes that the universe can be treated as if it
was isotropic after assuming
$B^{2}<3m_{pl}^{2}/\sqrt{N}$ \cite{Golovnev:2008cf}. An isotropic universe is implied in the following discussion.

The ``N-flation" vector field scenario has such an action
 \be
 \label{action}
 S=\int d^{4}x \sqrt{-g}[\frac{R}{16 \pi
 G}-N(\pm\frac{1}{4}F_{\mu\nu}F^{\mu\nu}+V(B^{2}))].
 \ee
With an additional function $F=\pm B^{2}/12$, the energy density, pressure and the equations of
motion  under these
assumptions can be simplified as
 \begin{eqnarray}
 \label{rhon}
 &&\rho=N\left(\pm\frac{1}{2}\dot{B}_{i}^{2}+V(B^{2})+6H(\dot{F}+H F)\right),\\
  \label{pn}
&& p=N\left(\pm\frac{1}{2}\dot{B}_{i}^{2}-V(B^{2})-2\ddot{F}-4H\dot{F}-2F(2\dot{H}+3H^{2})\right),\\
 &&\label{me}
  \ddot{B}_{i}+3H\dot{B}_{i}\pm6F'_{i}(2H^{2}+\dot{H})\pm V'_{i}=0.
  \end{eqnarray}
The above  forms in  ``N-flation" are just the same as those in the
``cosmic triad" scenario.  The energy density in Eqs. (\ref{rhos})
and (\ref{rhon})and the pressure in Eqs. (\ref{ps}) and (\ref{pn})
are different only in the coefficients which is characterized by the
number of the fields. Therefore in the following  we will only use
Eqs. (\ref{rhon}), (\ref{pn}) and (\ref{me}). The $N=3$ case is
regarded as the ``cosmic triad'' vector field scenario and the large
$N$ case is regarded as the ``N-flation" vector field scenario. And
based on these similarities, we will extend  minimal coupling to
non-minimal coupling for vector fields in the following discussions.

\subsection{Non-Minimal Coupling Vector Field Case}\label{subsec3}
In the vector field scenario, the non-minimal coupling term is used
to satisfy the slow-roll conditions. Without non-minimal coupling,
the vector field could only  be used as curvaton
\cite{Koivisto:2009sd,Dimopoulos:2006ms,Dimopoulos:2008rf,Dimopoulos:2008yv}.
To give a complete examination of the rolling, the possible
non-minimal coupling between the vector field and  gravity is also
included. Let us start from such an action
 \be
 \label{actionnon}
 S=\int d^{4}x \sqrt{-g}\left[\frac{R}{16 \pi G}+N\left(\pm\frac{1}{4}F_{\mu\nu}F^{\mu\nu}-V(A^{2})-f(A^{2})R\right)\right],
 \ee
 where the function $f(A^{2})$  shows the non-minimal coupling effect, and $A^{2}$ also can be
 rewritten as $B^{2}$ by Eq. (\ref{B}).
 Under ``cosmic triad" and ``$N$-flation" vector field assumptions,
the Friedmann and Raychaudhuri equations remain the same, but the
energy density, pressure and the equations of motion  are modified
to
\begin{eqnarray}
 &&\rho\simeq N\left(\pm\frac{1}{2}(\dot{B}_{i}+H B_{i})^{2}+V+6H(\dot{f}+H f)\right),\\
 &&p\simeq N\left(\pm\frac{1}{6}(\dot{B}_{i}+H
 B_{i})^{2}-V+\frac{V'_{i}B_{i}}{3}-2\ddot{f}-4H\dot{f}-2f(2\dot{H}+3H^{2})\right),\\
 &&\ddot{B}_{i}+3H\dot{B}_{i}\pm
 V'_{i}+(2H^{2}+\dot{H})B_{i}+6f'_{i}(\dot{H}+2H^{2})=0,
\end{eqnarray}
 where the non-minimal coupling parameter
 $f$ is the additional variable.
After  redefining the function $F=f(B^{2})\pm B^{2}/12$,  the above
equations can be reexpressed in a definite form
\begin{eqnarray}
 \label{rhonon}
 & &\rho\simeq N\left(\pm\frac{1}{2}\dot{B}_{i}^{2}+V+6H(\dot{F}+H F)\right),\\
 \label{pnon}
& & p\simeq N\left(\pm\frac{1}{2}\dot{B}_{i}^{2}-V-2\ddot{F}-4H\dot{F}
-2F(2\dot{H}+3H^{2})\right),\\
 \label{mnon}
&& \ddot{B}_{i}+3H\dot{B}_{i}+6F'_{i}(\dot{H}+2H^{2})\pm V'_{i}=0,
 \end{eqnarray}
which can be analyzed in the same as the minimal coupling case. The
latter   corresponds to  $f=0$ ($F=\pm B^{2}/12$). For convenience,
in the following section we first  discuss the non-minimally coupled
vector field case, then apply the results to the minimal coupling
case.

Especially, the $F=0$ case  corresponds  to the vector field
inflation discussed in Ref. \cite{Golovnev:2008cf}; furthermore
according to the equations
\begin{eqnarray}
 &&\rho\simeq N(\frac{1}{2}\dot{B}_{i}^{2}+V),\\
 && p\simeq N(\frac{1}{2}\dot{B}_{i}^{2}-V),
\end{eqnarray}
 the energy
density and the pressure are nearly the same as those in the
minimally coupled scalar fields.
 And as Ref. \cite{Golovnev:2008cf} argued, the slow-roll conditions
can be realized in this non-minimal coupling case.  The arguments
above indicate that  the behaviors of the vector
 fields  in  minimal  and   non-minimal coupling case  are
 totally different from those of the scalar fields. So it is necessary to investigate the vector field slow-roll
 conditions  specially.

\section{Slow-Roll Conditions in the Non-minimally Coupled Vector Field}\label{3}
The reason, that inflation needs slow-roll conditions, is  the
Hubble parameter $H$ should be nearly  constant and the
 universe should be in  de-Sitter phase for a long period of time.
 Combined with Eqs. (\ref{rhonon}), (\ref{pnon}) and (\ref{mnon}), by defining a function $\Omega=1-2F/m_{pl}^{2}$,
 Friedmann and Raychaudhuri equations for non-minimally coupled vector field
 become
  \begin{eqnarray}
  &&H^{2}\Omega+H\dot{\Omega}=\frac{N}{3m_{pl}^{2}}(\frac{1}{2}\dot{B}^{2}+V),\\
 \label{omega}
 &&\ddot{\Omega}-H\dot{\Omega}+2\dot{H}\Omega=-\frac{N\dot{B}^{2}}{3m_{pl}^{2}},
 \end{eqnarray}
and the equations of motion (\ref{me1}) could be rewritten as
 \be
 \ddot{B}_{i}+3H\dot{B}_{i}\pm
 V'_{i}\mp3m_{pl}^{2}\Omega'_{i}(\dot{H}+2H^{2})=0,
 \ee
where $m_{pl}^{-2}=8\pi G$ is the planck mass.  A long period of
inflation requires the potential dominates the evolution of the
universe, and the slow varying of the field means the acceleration
of the field should be neglected. The conditions $\dot{B}_{i}^{2}\ll
V$, $ |\dot{\Omega}|\ll H\Omega$, $\ddot{B}_{i}\ll H|\dot{B}_{i}|$
and $|\ddot{B}_{i}|\ll|V'_{i}|$ could reduce to
\begin{eqnarray}
 \label{H2}
 &&H^{2}\Omega\simeq\frac{NV}{3m_{pl}^{2}},\\
\label{mes}
 &&3H\dot{B}_{i}\simeq -\widetilde{V_{i}}',
 \end{eqnarray}
 where
$\widetilde{V_{i}}'=\pm
V'_{i}\mp6m_{pl}^{2}\Omega_{i}'H^{2}\mp3m_{pl}^{2}\Omega'_{i}\dot{H}$,
$\widetilde{V_{i}}$ could be regarded as the effective potential.
And we could define three parameter related to slow-roll process
\begin{eqnarray}
 \label{sl1}
 &&\epsilon\equiv\frac{m_{pl}^{2}\Omega \widetilde{V_{i}}'^{2}}{2NV^{2}},\\
\label{sl2}
 &&\eta\equiv\frac{m_{pl}^{2}\Omega \widetilde{V}''}{NV},\\
 \label{sl3}
 &&\delta\equiv\frac{m_{pl}^{2}\Omega_{i}'\widetilde{V_{i}}'}{NV}=\frac{-2F_{i}'\widetilde{V_{i}}'}{V}
\end{eqnarray}
for preparation. And for theoretical consistency, we need to know
the form of slow-roll conditions in detail that satisfy the
equations $\dot{B}_{i}^{2}\ll V$, $ |\dot{\Omega}|\ll H\Omega$,
$\ddot{B}_{i}\ll H|\dot{B}_{i}|$ and $|\ddot{B}_{i}|\ll|V'_{i}|$.
Varying Eq. (\ref{mes}) and using Eq. (\ref{H2}), we can get
 \begin{eqnarray}
 &&\frac{\dot{B}_{i}^{2}}{V}\simeq\frac{
 m_{pl}^{2}\Omega\widetilde{V}_{i}'^{2}}{3NV^{2}}=\frac{2}{3}\epsilon,\\
 &&\frac{\dot{\Omega}}{H\Omega}\simeq-\frac{m_{pl}^{2}\Omega_{i}'\widetilde{V}_{i}'}{NV}=-\delta,\\
 &&\frac{\ddot{B}_{i}}{H\dot{B}_{i}}\simeq-\frac{\dot{H}}{H^{2}}-\frac{m_{pl}^{2}\Omega
 \widetilde{V}''}{NV}=-\frac{\dot{H}}{H^{2}}-\eta,\\
&& \frac{\ddot{B}}{V'_{i}}\simeq\frac{\dot{H}}{3H^{2}}\frac{\widetilde{V}_{i}'}{V'_{i}}+\frac{m_{pl}^{2}\Omega
 \widetilde{V}''}{3NV}\frac{\widetilde{V}_{i}'}{V'_{i}}=\frac{\widetilde{V}_{i}'}{3V'_{i}}(\frac{\dot{H}}{H^{2}}+\eta),
 \end{eqnarray}
where $''=d^{2}/dB_{i}^{2}=d^{2}/dB^{2}$. By requiring
$|\epsilon|\ll1$ and $|\delta|\ll1$, $\dot{B}_{i}^{2}\ll V$ and $
|\dot{\Omega}|\ll H\Omega$ could be satisfied. And if we neglect
$\dot{\Omega}$ based on the definition of $\delta$,  Eq.
(\ref{omega}) can be rewritten as
$-(H\Omega)\dot{}+3\dot{H}\Omega=-N\dot{B}_{i}^{2}/3m_{pl}^{2}$.
Combining it with Eq. (\ref{H2}), we  obtain
 \be
 \label{dotH}
 \frac{\dot{H}}{H^{2}}=\frac{\dot{\Omega}}{2H\Omega}-\frac{m_{pl}^{2}\Omega
\widetilde{V}'^{2}}{6NV^{2}}=-\frac{\delta}{2}-\frac{\epsilon}{3},
  \ee
and we turns into $\dot{H}/H^{2}\ll1$ in the condition
$|\epsilon|\ll1$ and $|\delta|\ll1$. Moreover, adding the condition
$|\eta|\ll1$, we can get $\ddot{B}_{i}\ll H|\dot{B}_{i}|$ and
$|\ddot{B}_{i}|\ll|V'_{i}|$. In brief,
$|\epsilon|,|\delta|,|\eta|\ll1$ could be given out as slow-roll
conditions ground on the above discussions. Compared to the
minimally coupled scalar field case with only two slow-roll
parameters ($\epsilon$ and $\eta$), the additional slow-roll
parameter $\delta$ in the vector field scenario shows a constraint
on the function $\Omega$, in other word, on the function $F$, which
is not zero but $\pm B^{2}/2$ even in the minimally coupled vector
field scenario.

When $F=0$,   an special example of non-minimal coupling case
presented in Tab.(\ref{tab2}),
 we can get $\Omega=1$ and $\widetilde{V}_{i}'\simeq V'_{i}$.
And now the definition gives the parameter $\delta=0$ directly, so
$\delta$ can be ignored.  Furthermore, by putting  $\Omega=1$ and
$\widetilde{V}_{i}'\simeq V'_{i}$ into slow-roll conditions, we can
get $\epsilon=m_{pl}^{2}V_{i}'^{2}/2V^{2}$ and
$\eta=m_{pl}^{2}V''/V$ which are the same as the definitions of the
standard slow-roll parameters in the minimally coupled scalar
fields. It indicated that this special case may have the same good
property as the scalar field scenario.


 \begin{table}[ht]
\caption{Two cases.}
\begin{center}
  \begin{tabular}{l@{\hspace{2mm}}l@{\hspace{2mm}}l@{\hspace{2mm}}l@{\hspace{2mm}}l@{\hspace{2mm}}l@{\hspace{2mm}}cc}

    \hline
&Variable& Minimal coupling case & An special example\cite{Golovnev:2008cf}\\
\hline
 &$f(B^{2})$&$f=0$&$f=\mp B^{2}/2$\\
  &$F=\pm B^{2}/2+f(B^{2})$&$F=\pm B^{2}/2$&$F=0$\\
\hline
  \hline
  \end{tabular}
\end{center}
 \label{tab2}
\end{table}

\section{Rapid-Roll Conditions in the Non-minimally Coupled Vector Field}\label{4}
As presented in the introduction,  getting a de-Sitter phase is
critical in the early and late period of acceleration, which
demands $\dot{H}/H^{2}\ll1$. As Eq. (\ref{dotH}) noticed,
$\dot{H}/H^{2}\ll1$ only requires $\epsilon, \delta\ll1$. Hence,  if
we only ask for a de-Sitter phase without considering how long it
would last, the slow-roll conditions could be relaxed, especially
the condition related to $\eta$ parameter. That is the motivation of
 rapid-roll inflation
\cite{Kofman:2007tr,Chiba:2008ia,Chiba:2008rp}, which could be
regarded as a new model of  fast-roll inflation \cite{Linde:2001ae}.
The rapid-roll inflation  can lead a period of acceleration
(de-Sitter phase) as well, but requires looser conditions compared
to slow-roll inflation.

However, the scalar field  rapid-roll inflation is a novel type of
inflation with a non-minimal coupling term in which inflation could
occur without slow-rolling field \cite{Chiba:2008ia}. Here, because
the similarity between the equations in the minimally  and  non-minimally coupled scalar field
scenario, the minimally coupled vector field could induce rapid-roll
inflation as well. We will define the vector field rapid-roll
inflation by following Ref. \cite{Chiba:2008ia}, which starts with the
Friedmann Equation and the equations of motion
\begin{eqnarray}
&& H^{2}=\frac{N}{3m_{pl}^{2}}\left(\pm\frac{1}{2}\dot{B}_{i}^{2}+V+6H(\dot{F}+H F)\right),\\
 &&\ddot{B}_{i}+3H\dot{B}_{i}\pm V'_{i}\pm 6F'_{i}(\dot{H}+2H^{2})=0.
 \end{eqnarray}
After defining a new available viable $\vartheta_{i}=\dot{B}_{i}\pm
6HF'_{i}$, the above equations can be rewritten as
 \begin{eqnarray}
 &&\label{rrH}
 H^{2}=\frac{N}{3m_{pl}^{2}}\left(\pm\frac{1}{2}\vartheta_{i}^{2}+V+6H(\dot{F}-3F'^{2})\right),\\
 \label{rrm}
 &&\dot{\vartheta_{i}}+2H\vartheta_{i}\pm V'_{i}+(1\mp6F'')H\dot{B}_{i}=0.
 \end{eqnarray}
It is still need to assume that the potentials of vector fields
 dominate over the kinetic energy term in the rapid-roll
inflation. Meanwhile,  we provides a dimensionless parameter $c$ to
loose the rolling of vector fields. That means Eqs. (\ref{rrH}) and
 (\ref{rrm}) should be simplified to
\begin{eqnarray}
 \label{H2R}
 &&H^{2}\simeq \frac{NV}{3m_{pl}^{2}},\\
 \label{mer}
 &&(c+2)H\vartheta_{i}\simeq-V'_{i}.
  \end{eqnarray}
And we could define three parameters
 as following for
preparation:
 \begin{eqnarray}
 \label{c1}&&\epsilon_{c}\equiv\pm\frac{m_{pl}^{2}V_{i}'^{2}}{2NV^{2}}+\frac{2N(c+2)^{2}}{3m_{pl}^{2}}(F-3F_{i}'^{2}),\\
 &&\eta_{c}\equiv\frac{m_{pl}^{2}V''}{NV}+\frac{2(c+2)F_{i}'V''}{V_{i}'}+\frac{c(c+2)}{3}-\frac{c+2}{3}(1\mp6F'')
 \mp\frac{2N(c+2)^{2}}{3}\frac{(1\mp6F'')F_{i}'V}{m_{pl}^{2}V_{i}'},\\
 \label{c2}&&\delta_{c}\equiv\frac{m_{pl}^{2}V_{i}'^{2}}{2N(c+2)V^{2}}\pm\frac{F_{i}'V_{i}'}{V}.
\end{eqnarray}
 To  satisfy  Eqs. (\ref{H2R}) and (\ref{mer}), we
needs the following inequalities
 \begin{eqnarray}
 \label{rl1}
 &&\frac{\pm\vartheta_{i}^{2}/2+6H^{2}(F-3F_{i}'^{2})}{V}\simeq\frac{\pm3m_{pl}^{2}V_{i}'^{2}}{2N(c+2)^{2}V^{2}}+\frac{2N}{m_{pl}^{2}}(F-3F_{i}'^{2})
 =\frac{3}{(c+2)^{2}}\epsilon_{c}\ll1,\\
 &&\label{rl2}
  \nonumber \frac{\dot{\vartheta_{i}}-cH\vartheta_{i}+H(1\mp6F'')(\vartheta_{i}\mp6HF'_{i})}{(c+2)H\vartheta_{i}}\\
  \nonumber &  &
  \simeq -\frac{\dot{H}}{(c+2)H^{2}}-\frac{3m_{pl}^{2}V''}{NV(c+2)^{2}}-\frac{6F_{i}'V''}{(c+2)V'_{i}}
  -\frac{c}{c+2}
  +\frac{1-6F''}{c+2}\pm\frac{2(1\mp6F'')F'_{i}V}{m_{pl}^{2}V'_{i}}\\
  &&=-\frac{\dot{H}}{(c+2)H^{2}}-\frac{3}{(c+2)^{2}}\eta_{c}\ll1.
\end{eqnarray}
Eq. (\ref{rl2}) require
 $\dot{H}/H^{2}\ll1$, the de-Sitter phase as well. So, for  the consideration of consistence,
by using Eq.
 (\ref{H2R}) we get
 \be
 \label{rl3}
 \left|\frac{\dot{H}}{H^{2}}\right|=\left|\frac{-3m_{pl}^{2}V_{i}'^{2}}{2N(c+2)V^{2}}\mp\frac{3V'_{i}F'_{i}}{V}\right|=\left|-3\delta_{c}\right|.
\ee
 The above equation notes after requiring  $|\epsilon_{c}|,|\eta_{c}|,|\delta_{c}|\ll1$,
 Eqs. (\ref{H2R}) and (\ref{mer}) could be consistently satisfied.

 The value of $c$ is determined by $\eta_{c}$, as the rapid-roll conditions require $\eta_{c}\simeq 0$,
 we could
get
 \be
 \label{cc}
 c=-2-\frac{x}{2y}\pm^{*}\sqrt{(\frac{x}{2y})^{2}-\frac{m_{pl}^{2}V''}{y N V}},
 \ee
where the sign $``\pm^{*}" $ suggests two different solutions,
$x=2F'_{i}V''/V'_{i}-2/3-(1\mp6F'')/3$ and
 $y=\mp2NF'_{i}V(1\mp6F'')/3m_{pl}^{2}V'_{i}+1/3$.

Take $F=0$ for example, the rapid-roll conditions can be reduced to
 \begin{eqnarray}
 &&\epsilon_{c}=\pm\frac{m_{pl}^{2}V_{i}'^{2}}{2NV^{2}}\ll1,\\
 &&\eta_{c}=\frac{m_{pl}^{2}V''}{NV}+\frac{(c-1)(c+2)}{3}\ll1,\\
 &&\delta_{c}=\frac{m_{pl}^{2}V_{i}'^{2}}{2N(c+2)V^{2}}\ll1.
 \end{eqnarray}
If  $\eta_{c}\simeq0$,
 \be
 c=-\frac{1}{2}\pm^{*}\sqrt{\frac{9}{4}-\frac{3m_{pl}^{2}V''}{ N V}}.
  \ee
The differences between  $\epsilon_{c}$ and $\delta_{c}$ are in the
coefficients, so either is enough. Particularly speaking, the number
of the slow-roll parameters could be reduced to two. And when
$V''\ll V$, $c=1$, the form of $\eta_{c}$ ($\eta_{c}\simeq
m_{pl}^{2}V''/V$)  is similar to the $\eta$ parameter definition in
the scalar field. Therefore, when $V''\ll V$,  the rapid-roll vector
field inflation coincides with the slow-roll vector field case.

\section{Minimally Coupled Vector Field}\label{sec5}
\subsection{Slow-Roll and Rapid-Roll Conditions}
As Tab.(\ref{tab2}) notes, the minimally coupled  vector field case
corresponds to $F=\pm B^{2}/12$. From the minimal coupling
view, we can get $\Omega=1\mp B^{2}/(6m_{pl}^{2})$, $\Omega'_{i}=\mp
B_{i}/(3m_{pl}^{2})$ and $\widetilde{V}_{i}'\simeq \pm
V'_{i}+2H^{2}B_{i}$. Using these conditions,  the slow-roll
conditions could be obtained
\begin{eqnarray}
 \label{sls1}
 &&\epsilon=\frac{\Omega}{2N}(\frac{m_{pl}V'_{i}}{V}+\frac{2NB_{i}}{3m_{pl}} \frac{1}{1-B^{2}/6m_{pl}^{2}})^{2}\ll1,\\
 &&\label{sls2}
 \eta=\frac{V''\mp2H^{2}}{3H^{2}}\ll1,\\
 &&\label{sls3}
 \delta=-\frac{V'_{i}B_{i}}{3NV}\mp\frac{2B^{2}}{9m_{pl}^{2}}\frac{1}{(1-B^{2}/6m_{pl}^{2})}\ll1.
\end{eqnarray}
 In small-field potential
model which means $B^{2}\ll m_{pl}^{2}$, the second term in the slow-roll parameters $\epsilon,\delta$
would be much smaller than $1$. But for  large-field potential model
where $B^{2}\geq m_{pl}^{2}$, we could not get such result.
According to the above argument, it seems easier to satisfy
slow-roll conditions
 in the small-field potential model.
And Eq. (\ref{sls2}) gives $V''\simeq\mp2H^{2}$ which is hard to
achieve, at least needing a certain level of tuning.

As for the rapid-roll conditions in the minimal coupling case, we
can obtain
\begin{eqnarray}
 &&\label{slr1}
 \epsilon_{c}=\pm\frac{m_{pl}^{2}V_{i}'^{2}}{2NV^{2}}\ll1,\\
 &&\label{slr2}
 \eta_{c}=\frac{m_{pl}^{2}V''}{NV}\pm \frac{(c+2)B_{i}
 V''}{3V'_{i}}+\frac{c(c+2)}{3}\ll1,\\
  &&\label{slr3}
 \delta_{c}=\frac{\mp\epsilon_{c}}{c+2}\pm\frac{B_{i} V'_{i}}{6V}\ll1.
\end{eqnarray}
From Eq. (\ref{slr1}), we can see that  $\epsilon_{c}\ll1$ can
always be satisfied as far as $N$ is large enough. And  Eq.
(\ref{slr3}) requires $B_{i} V'_{i}/(6V)\ll1$, which could be
satisfied easier in the small-field model than the large-field model
as in the situation of slow-roll conditions. Then $\eta_{c}\simeq0$
requires
 \be
 c=-(1+\frac{V''B_{i}}{2V'_{i}})\pm^{*}\sqrt{(1-\frac{V''B_{i}}{2V'_{i}})^{2}-3\frac{m_{pl}^{2}V''}{NV}}.
 \ee

\subsection{Example}\label{sec3}
From the view of gravitational waves, it is interesting that Ref.
\cite{Kobayashi:2009hj,Golovnev:2008hv} pointed out only the small
field models are feasible. In the following discussions, we will
take two examples of potential to discuss the feasible models
explicitly from the view of slow-rolling and rapid-rolling. One is
the large-field model (chaotic potential $V=m^{2}B^{2}/2$) and the
other is the small-field model ($V=V_{0}-m^{2}B^{2}/2$ for the
positive kinetic energy term case and $V=V_{0}+m^{2}B^{2}/2$ for the
negative kinetic term case). More definite expressions are present
in Tab.(\ref{tab3}).

In the large-field potential $V=m^{2}B^{2}/2$, for the slow-roll
conditions, it is impossible for the parameter $\eta$ to be
satisfied in the positive kinetic energy case, and in the negative
kinetic energy case only when $m^{2}\simeq2H^{2}$ it could be
satisfied. But when $B^{2}/m_{pl}^{2}\gg1$,  we can find that the
first term  in the brackets of Eq. (\ref{sls3}) is so large that it
is not suitable for the slow-roll conditions. When
$B^{2}/m_{pl}^{2}\simeq1$ this problem is more serious since both
Eqs. (\ref{sls1}) and (\ref{sls3}) could not be satisfied simultaneously. So the
chaotic potential is not proper for the slow-roll conditions. With
reference to the rapid-roll conditions, the second term in Eq.
(\ref{slr3}) which is $1/3$  can exclude the model.

For the small-field potential,  as $B^{2}\ll m_{pl}^{2}$ and
$m^{2}B^{2}/2\ll V_{0}$, the slow-roll conditions that
$\epsilon,\delta\ll1$ can be satisfied. But for the $\eta$
parameter, the problem   still exists that $m^{2}$ should be at
order of ${\cal O}(2H^{2})$. Fortunately in the rapid-roll
conditions, as $m^{2}B^{2}/2\ll V$, $\epsilon_{c}$, $\eta_{c}$ and
$\delta_{c}$ can be fully satisfied with $c=-1$.


 \begin{table}[ht]
\caption{Small-field and large-field potential for slow-roll and
rapid-roll inflation.}
\begin{center}
  \begin{tabular}{l@{\hspace{2mm}}l@{\hspace{2mm}}l@{\hspace{2mm}}l@{\hspace{2mm}}l@{\hspace{2mm}}l@{\hspace{2mm}}cc}

    \hline
&Potentials&Expression& Slow-roll& Rapid-roll \\
\hline
  &Small-field&$V=V_{0}-m^{2}B^{2}/2$ (Positive kinetic energy) & $m^{2}\sim
{\cal O}(2H^{2})$& Proper\\
&&  $V=V_{0}+m^{2}B^{2}/2$ (Negative kinetic energy)& $m^{2}\sim
{\cal O}(2H^{2})$&Proper\\
  &Large-field & $V=m^{2}B^{2}/2$(Chaotic potential)&Not proper& Not proper\\
\hline
  \hline
  \end{tabular}
\end{center}
 \label{tab3}
\end{table}

\section{Summary}\label{sec3}
In the above discussions, we give out the exact forms of the
slow-roll and rapid-roll conditions in the space-like vector field
inflation. After the assumptions in ``cosmic triad" and ``N-flation"
vector field scenarios being set up, we could see the  forms of the
minimally coupled vector fields are similar to the non-minimally
coupled scalar fields. Because  the increasing number of the
slow-roll parameters (there is an additional parameter $\delta$) and
the  fine-tuning model parameters, it is natural that the slow-roll
conditions in the vector field scenarios are stricter than
 the scalar field scenario. When $F=0$, an special example in the non-minimal coupling case, the
slow-roll conditions in  the vector field are nearly the same as the
minimally coupled scalar field. However, in the minimally coupled
vector field case,  the slow-roll conditions  is much more probable
to realize in small-field models rather than in large-field models.
And the rapid-roll inflation, as a new model of fast-roll inflation,
requires much looser conditions compared with slow-roll inflation,
particularly the constraint on $\eta$ as the examples show.
Nevertheless, the positive and negative kinetic energy cases in this
model have a lot of similar behaviors. This subject should be
further investigated especially from the physical model building
aspect.


\section*{Acknowledgements}
YZ thanks Rong-gen Cai,  Hui Li, Bin Hu, Ya-wen Sun, Xian Gao and
Jian-feng Cheng for useful discussions, and the referee for many
helpful suggestions. This work was supported in part by a grant from
Chinese Academy of Sciences, grants from NSFC with No. 10325525 and
No. 90403029.



\end{document}